\documentclass[11pt,twoside]{article}

\usepackage{asp2010}
\usepackage{graphicx}

\def\msun{\,{\rm M_\odot}}

\newenvironment{itemize_estret}{
\begin{itemize}
  \setlength{\itemsep}{1pt}
  \setlength{\parskip}{0pt}
  \setlength{\parsep}{0pt}
}{\end{itemize}}

\resetcounters 

\begin{document}

\resetcounters

\title{Detecting massive black hole binaries and unveiling their cosmic history with gravitational wave observations}
 \author{Alberto Sesana$^1$ \affil{$ö1$ Max-Planck-Institut f{\"u}r Gravitationsphysik, Albert Einstein Institut, Am M\"uhlenberg 1, 14476 Golm, Germany}}
 
\begin{abstract} 
Space based gravitational wave astronomy will open a completely new window on the Universe and massive black holes binaries are expected to be among the primary actors on this upcoming stage. The New Gravitational-wave Observatory (NGO) is a space interferometer proposal derived from the former Laser Interferometer Space Antenna (LISA) concept. We describe here its capabilities of observing massive black hole binaries throughout the Universe, measuring their relevant parameters (masses, spins, distance to the observer) to high precision. The statistical properties of the population of detected systems can be used to constrain the massive black hole cosmic history, providing deep insights into the faint, high redshift Universe.
\end{abstract}

\section{Introduction}
Today, massive black holes (MBHs) are ubiquitous in the nuclei of nearby galaxies ~\citep{mago98}. However, our current knowledge of the MBH population is limited to a small fraction of objects: either those that are active, or those in our neighborhood, where stellar- and gas-dynamical measurements are possible. In $\Lambda$CDM cosmologies, structure formation proceeds in a hierarchical fashion \citep{wr78}, in which massive galaxies are the result of several merging events involving smaller building blocks. As a consequence, the MBHs we see in today's galaxies are expected to be the natural end-product of a complex evolutionary path, in which black holes seeded in proto-galaxies at high redshift grow through cosmic history via a sequence of MBH-MBH mergers and accretion episodes \citep{kh00,vhm03}. In this framework, a large number of MBH binaries naturally form following the frequent galaxy mergers. MBH binaries are the loudest gravitational wave (GW) sources in the Universe \citep[e.g.,][]{hughes02}: space based GW interferometers will open a completely new window on the cosmos, revealing the population of these electromagnetically invisible systems.

Over the last two decades ESA and NASA developed the concept of  a space-based GW observatory that would explore the low frequency GW sky, in the frequency band $10^{-4}-1$ Hz, suitable to the detection of coalescing MBH binaries throughout the Universe: the Laser Interferometer Space Antenna \citep[LISA,][]{bender98}. The collaboration between the two agencies ended in early 2011 for programmatic and budgetary reasons. During that year, a team of scientists worked to a modification of the LISA concept, to be considered by ESA alone for launch in 2022 within the Cosmic Vision program. This effort resulted in the proposal of the New Gravitational-wave Observatory \citep[NGO,][]{amaro12}; a smaller observatory inspired by the LISA design, featuring a triangular spacecraft formation with 1Mkm arms.

In this paper we report on the MBH binary science that a spaced based detector like NGO can deliver. Despite the inevitable loss in sensitivity with respect to LISA, NGO can still detect GWs from coalescing binaries with total masses in the range $10^3-10^6\,\msun$ out to $z\sim 20$. Masses, spins and distances (encoded in the GW waveform) of individual coalescing MBHs will be measured with unprecedented precision. Moreover, observations of multiple mergers can be combined to extract useful astrophysical information about their formation and evolution through cosmic history. In fact, the analysis of the overall distribution of the coalescing MBH binary population parameters offers the unique possibility of discriminating between different black hole seed formation mechanisms and accretion modes \citep{plowman11,sesana11}. This, in turn, can be used to place strong constrains on the structure formation process in a redshift and mass range inaccessible to electromagnetic observations. 

This article is organized as follows. In Sec. 2 we describe the cosmological models for MBH formation and evolution employed as a testbed of the NGO capabilities. In Sec. 3, we discuss the NGO observations of individual MBH binaries, describing the performances of the detector and assessing is capabilities in extracting the parameters of the sources. In Sec. 4 we turn to consider the whole population of detected binaries: we show that the distribution encodes valuable astrophysical information that allows to constrain the high redshift formation and early evolution of these systems.  We summarize and conclude in Sec. 5.    

\section{Massive black hole binaries in the framework of galaxy assembly}
%In the currently favored $\Lambda$CDM framework, the MBHs we see in galaxies today are expected to be the natural end-product of a complex evolutionary path, in which black holes seeded in proto-galaxies at high redshift grow along the cosmic history through a sequence of mergers and accretion episodes \citep{kh00,vhm03}. Hierarchical models for MBH evolution, associating quasar activity to gas-fueled accretion following mergers between galaxies, have been successful in reproducing several properties of the observed Universe, such as the present day mass density of nuclear MBHs and the optical and X-ray luminosity functions of quasars \citep{vhm03,m07}. In these models, a large number of MBHBs naturally form: these are the targets of NGO.

The MBHs we see in galaxies today are expected to be the natural end-product of the hierarchical structure formation paradigm. In this picture, black holes seeded in proto-galaxies at high redshift grow along the cosmic history through a sequence of mergers and accretion episodes. However, the mechanism responsible for the formation of the first seed black holes that will evolve in the MBHs we see today is not well understood, and two distinctive families of models have became popular in the last decade. In the first family, seeds are light ($M\sim100\msun$, 'small seed' scenario), being the remnant of the first PopIII star explosions \citep{mr01}; in the second one, the 'large seed' scenario, already quite heavy ($M\sim10^4\msun$) seed black holes form by direct collapse of massive proto-galactic discs \citep[see, e.g.,][]{kbd04,bvr06}. Moreover, the details of their subsequent mass growth are poorly understood. Here we test two ideal distinct situations, in which all the MBHs efficiently accrete gas either in a coherent or in a chaotic fashion. In the first case MBHs accrete matter from long-lasting extended disks, rapidly aligning their spins with the disk angular momentum vector \citep{bardeen75}. The net result is an efficient MBH spin-up \citep{thorne74}. In the second case, the MBH is fueled by small clumps of matter coming from random orientations, keeping the MBH spin small \citep[chaotic accretion,][]{king05}. The two accretion scenarios readily result in a very different MBH spin distribution, but also predict a different pace of the MBH mass growth. Smaller spins imply smaller mass/radiation conversion efficiency, resulting in a faster mass growth that leaves an imprint on the mass distribution as a function of cosmic time. 

To assess the astrophysical impact of NGO, we combine the two seed black hole and accretion prescriptions introduced above to produce cosmological MBH evolution scenarios labeled as follows: 
\begin{itemize_estret}
\item SE: seeds have small (S) mass ($\sim100\msun$) and accretion is coherent, i.e. resulting in an efficient (E) spin-up:
\item SC: seeds are small but accretion is chaotic (C);
\item LE: seeds are large (L), $\sim10^4\msun$, and accretion in coherent;
\item LC: large seeds and chaotic accretion.
\end{itemize_estret}
The models are almost the same used in previous studies by the LISA Parameter Estimation Task Force \citep{arun09}. The only difference is that in the efficient accretion model, spins are not assumed to be perfectly aligned to the binary orbital angular momentum. The angles of misalignment relative to the orbit are drawn randomly in the range 0 to 20 degree, consistent with the finding of recent hydrodynamical simulations of binaries forming in wet mergers \citep{dotti10}. These models encompass a broad range of plausible massive black hole evolution scenarios, and we use them as a testbed for NGO capabilities in a fiducial astrophysical context.  

Any massive black hole binary, coalescing at redshift $z$, is characterised by the (rest frame) total mass $M=m_1+m_2$ (with $m_1$ and $m_2$ the mass of the primary and secondary black hole), mass ratio $q=m_2/m_1$, spin vectors ${\bf{J}_1}$ and ${\bf{J}_2}$; spin magnitudes are denoted by $a_1$ and $a_2$. Each model (SE, SC, LE, LC) predicts a peculiar distribution $N(M,q,z,a_1,a_2)$ of coalescing binaries. The orientations of the spins are drawn as described above for the efficient (E) accretion models, and completely random for the chaotic (C) accretion models. We generate several Monte Carlo realisations of each model starting from the $N(M,q,z,a_1,a_2)$ distribution and we create catalogues of coalescing MBH binaries randomising other source parameters (sky location, polarisation, inclination, initial phase, coalescence time) according to the appropriate distribution. NGO performances are tested against these fiducial catalogues of potentially detectable MBH binary coalescences.

% we sum up all the generated sources in a single ``average'' catalogue (we will consider models separately in the next section). 
%.Catalogues are generated by selecting $M$, $q$, $z$, $a_1$, $a_2$ according to the distribution predicted by the individual models, and by

\section{NGO observations of MBH binaries}

\subsection{Noise curve and waveform models}
\begin{figure}[!h]
\begin{center}
\includegraphics[scale = 0.4, angle = 0]{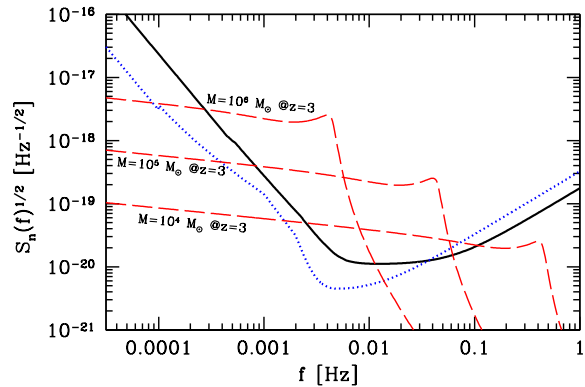}
\caption{\label{fig1} Sensitivity curves of NGO (solid black line) and LISA (dotted blue line) compared to the sky-polarization averaged strain produced by the inspiral+merger+ringdown of selected equal mass binaries (dashed red lines). The latter are generated using PhenomC non spinning waveforms.
}
\end{center}
\end{figure}

The NGO design inherits most of the LISA features, with two major differences: (i) the three satellites  maintain a near-equilateral triangular formation with an armlength $L=10^9$ m, five times smaller than LISA; and (ii) a mother-daughters configuration is adopted. The mother spacecraft serves as the "central hub" and houses two free-falling "test masses", defining the vertex of the "V" configuration. The other spacecrafts contain one test mass only, defining the two endpoints of the "V". Four laser links connect the test masses in a single-Michelson configuration. Point (i) causes the sensitivity curve to shift to higher frequencies with respect to LISA, while point (ii) prevents the instantaneous independent measurement of the two GW polarizations, somewhat limiting the GW signal reconstruction. Other technical differences related to the laser power and the collecting mirror size, also cause the noise floor level to be a factor of few worse than LISA. The sensitivities of the two instruments are compared in figure \ref{fig1}. 

%\vspace{0.5cm}

Massive black hole binaries sweep the NGO sensitivity window (figure \ref{fig1}) from the left to the right, and the detected signal includes all the three steps of their late evolution: inspiral, merger, and ring-down. While the inspiral and the ringdown can be treated analytically via Post Newtonian (PN) expansion of their binding energy and radiated flux \citep{blanchet06}, and black hole perturbation theory respectively, the merger  can only be described by a numerical solution of the Einstein equations \citep{pret05,campanelli06,baker06}. In recent years there has been a major effort in constructing accurate waveforms inclusive of all three phases. "Complete" waveforms can be designed by stitching together analytical PN waveforms for the early inspiral with a (semi)phenomenologically described merger and ring-down phase \citep[see][and references therein]{ohme12} calibrated against available numerical data.  In the following estimations we will mostly employ phenomenological waveforms constructed in frequency domain, as described in \citep{santamaria10}.  Self-consistent waveforms of this type (known as PhenomC waveforms) are available for non-spinning binaries and for binaries with aligned spins. In the case of binaries with misaligned spins, we use "hybrid" waveforms obtained by stitching precessing PN waveforms for the inspiral with PhenomC waveforms for the merger/ring-down. This stitching is performed by projecting the orbital angular momentum and individual spins onto the angular momentum of the distorted black hole after merger.

\subsection{Detector performances}

\begin{figure}[!h]
\plottwo{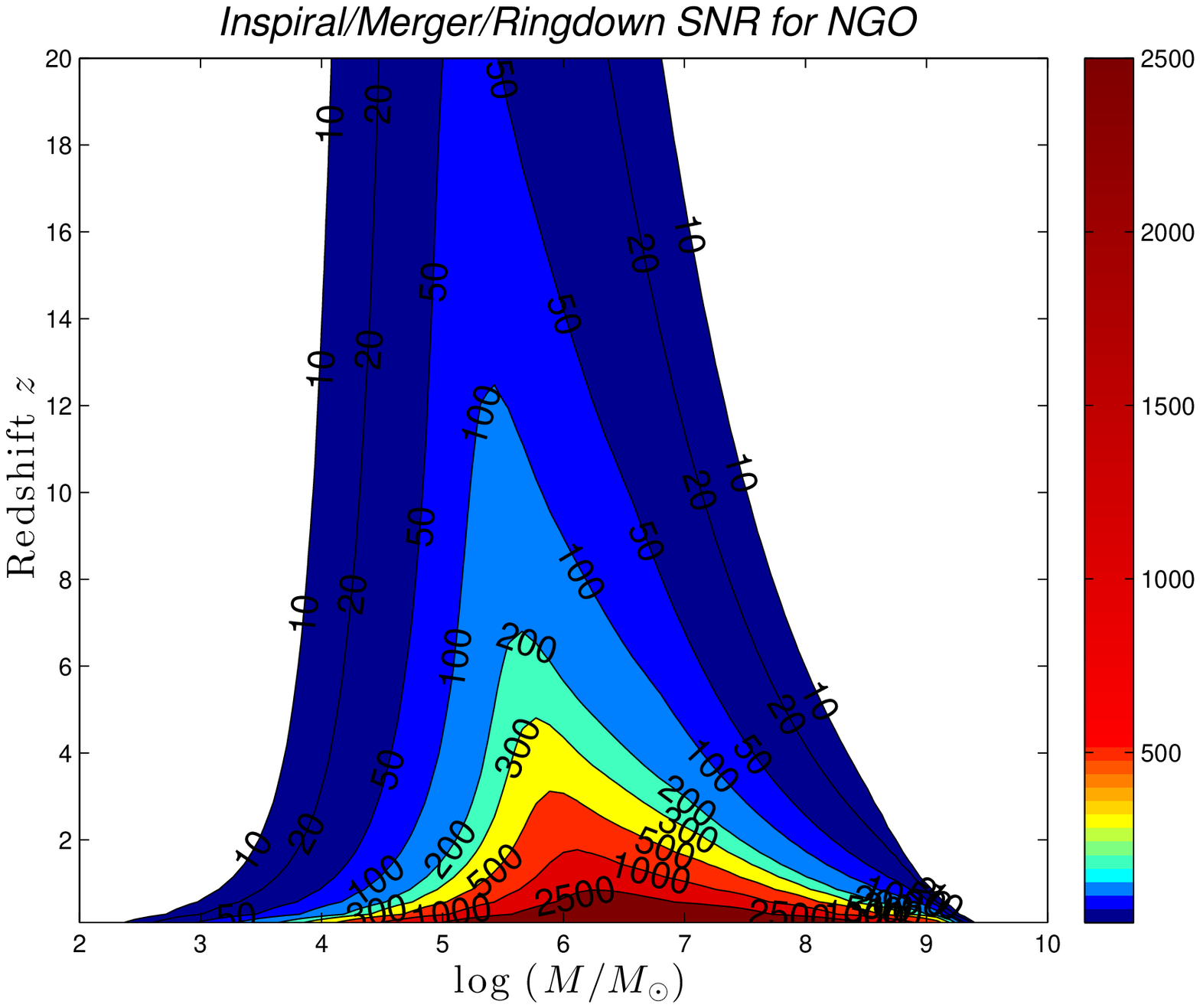}{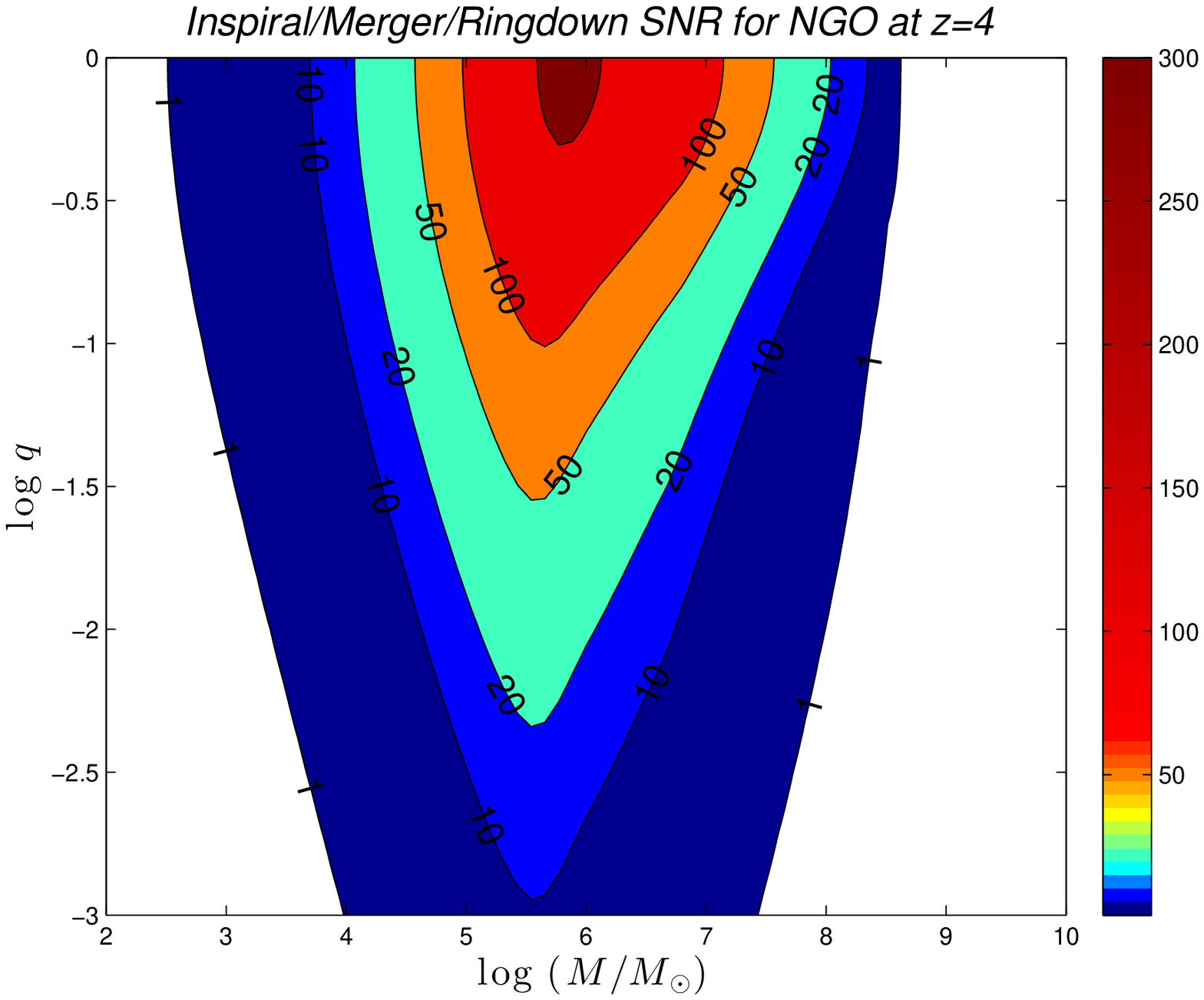}
\caption{\label{fig2} Left panel: constant-contour levels of the sky and polarisation angle-averaged SNR for equal mass non-spinning binaries as a function of their total mass $M$ and cosmological redshift $z$. Left panel: constant-contour levels of the sky and polarisation angle-averaged SNR for unequal mass non-spinning binaries placed at $z=4$ as a function of their total mass $M$ and of the mass ratio $q$. The total mass $M$ is measured in the rest frame of the source. 
%The SNR is computed using PhenomC
%    waveforms \citep{santamaria:2010PhRvD..82f4016S}, which are
%    inclusive of the three phases of black hole coalescence (in
%    jargon: inspiral, merger, and ring-down, as described in the
%    text).
}
\end{figure}
Given a waveform model, a first measure of the NGO performances is the signal-to-noise ratio (SNR) of a binary merger with parameters in the relevant astrophysical range. The left panel of figure \ref{fig2} shows NGO SNRs for equal mass, non-spinning coalescing binaries. Here we use PhenomC waveforms and we compute the SNR as a function of the rest-frame total binary mass $M$ and of the redshift $z$, averaging over all possible source sky locations and wave polarisation, assuming two-year observations. The plot highlights the exquisite capabilities of the instrument in covering almost all the mass-redshift parameter space relevant to MBH astrophysics. Conversely, our current MBH knowledge is bound to instrument flux limits, covering only the mass range $10^7-10^9\msun$ at $0\lesssim z\lesssim 7$. NGO will be able to detect the GWs emitted by sources with $M\sim10^4\msun$ at cosmological distances inaccessible to any other astrophysical probe. A binary with $10^4\lesssim M\lesssim 10^7\msun$ can be detected out to  $z\sim 20$ with a SNR $\ge 10$, making NGO optimal for a deep and extensive census of the MBH population in the Universe. In the right panel of figure \ref{fig2} we show constant-contour levels of the SNR expected from binaries with different mass ratios $q$ located at redshift $z=4$. Despite the natural SNR reduction that occurs with decreasing $q$, binaries in the mass range $10^5-10^7\msun$ with $q\lesssim\ensuremath{10^{-1}}$ can be detected with SNR $>20$  even at this redshift.

\subsection{Parameter estimation}
Detected waveforms carry information on the redshifted mass (the mass measured at the detector is $(1+z)$ times the mass at the source location) and on the spin of the individual black holes prior to coalescence. The measure of these parameters are of great importance in astrophysics. Masses of extragalactic MBHs are estimated with uncertainties ranging from $15\%$ to a factor $\approx2$ \citep[see,e. g.,][]{gebhardt00,vestergaard02}, depending on the technique used and the type of source. As far as spins are concerned, their measure is only indirect, and it is derived through modeling of the spectrum, or of the shape of emission lines, mainly by fitting the skewed relativistic K$\alpha$ iron line \citep[see][for a review]{miller07}. There are few notable examples, but uncertainties are still large. By contrast, spins leave a distinctive imprint in the waveform.

\begin{figure}[!h]
\begin{center}
\includegraphics[scale = 0.5, angle = 0]{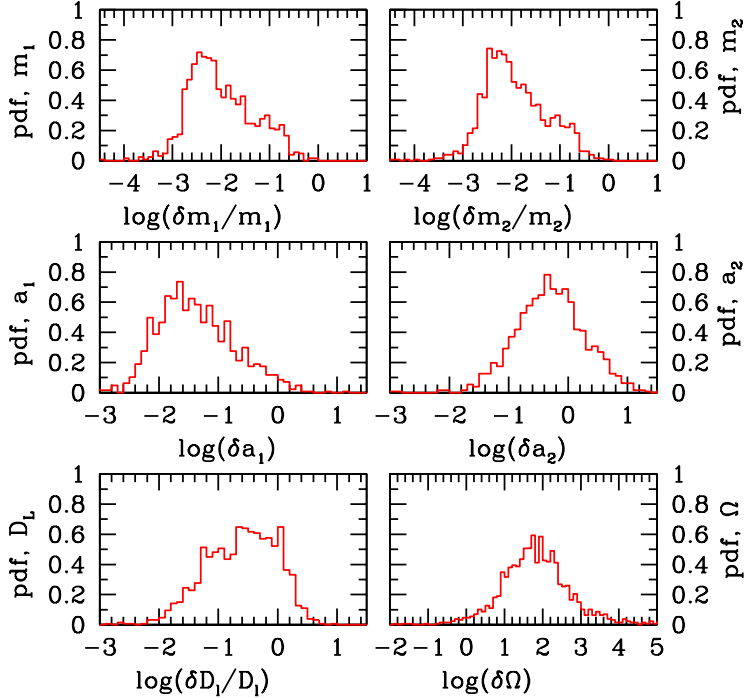}
\caption{\label{fig3}Parameter estimation accuracy evaluated on a source catalogue, obtained combining ten Monte Carlo realisations of the coalescing massive black hole binary population, predicted by the four SE-SC-LE-LC models. The top panels show the distributions of the fractional errors in the estimation of the redshifted masses of the primary (left) and secondary (right) black hole. The middle panels show the absolute error distributions on the measurement of the primary (left) and secondary (right) black hole spin, while the bottom panels show the fractional error distribution on the luminosity distance $D_L$, and the sky location accuracy $\Delta\Omega$ (in deg$^2$). 
%Errors are evaluated considering the full waveforms.
}
\end{center}
\end{figure}

Figure \ref{fig3} shows error distributions in the source parameter estimation, for events collected in a meta-catalogue of $\sim 1500$ sources. The meta-catalogue is the sum of four catalogues for the individual MBH evolution models (SE SC LE LC) containing 10 Montecarlo realizations of the Universe each. We used the "hybrid" waveforms described above, to evaluate uncertainties based on the Fisher information matrix (FIM) approximation. Individual redshifted masses can be measured with unprecedented precision, i.e. with an error of $0.1\%-1\%$, on both components. No other astrophysical tool has the capability of reaching a comparable accuracy on extragalactic MBHs. The spin of the primary hole can be measured with an exquisite accuracy, to a 0.01-0.1 absolute uncertainty. This precision mirrors the big imprint left by the primary MBH spin in the waveform. The measurement is more problematic for $a_2$ that can be either determined to an accuracy of 0.1, or remain completely undetermined, depending on the source mass ratio and spin amplitude. We emphasise that the spin measure is a neat, direct measurement, that does not involve complex, often degenerate, multi-parametric fits of high energy emission processes. The source luminosity distance error $D_L$ has a wide spread, usually ranging from being undetermined to a stunning few percent accuracy. Note that this is a direct measurement of the luminosity distance to the source, which, again, cannot be directly obtained (for cosmological objects) at any comparable accuracy level by any other astrophysical means. NGO is a full sky monitor, and the localisation of the source in the sky is also encoded in the waveform pattern. Sky location accuracy is typically estimated in the range 10-1000 square degrees.

\vspace{2.0cm}

\subsection{A note about the luminosity distance}
$D_L$ is usually undetermined for weak sources at high $z$, which are maybe the most interesting from a cosmological prospective. The distance information is encoded in the amplitude $A$ of the measured GW inspiral, which can be expressed in terms of the source parameters as: 
\begin{equation}
A \propto \frac{{\cal M}^{5/6}}{F\times D_L},
\end{equation}
where ${\cal M}$ is the chirp mass of the system, and $F$ is a known function of the source sky location, inclination and polarization angle. If we know ${\cal M}$ and $F$, we can infer $D_L$ from the direct measurement of $A$ (with an error of $\approx 1/$SNR). In the unlucky circumstance of rather weak signals, sky location, polarization angle and inclination are degenerate and poorly constrained, preventing a reliable estimation of $D_L$. In general, relative errors of 100\% on $D_L$ arise from three correlated reasons: (i) low SNR, where the linear approximation (assumed by the FIM) is not sufficient; (ii) partial or complete degeneracy of the parameters; (iii) highly non-Gaussian posterior distribution of some parameters (the FIM assumes Gaussian posteriors). 

\vspace{1.0cm}
\begin{figure}[!h]
\plottwo{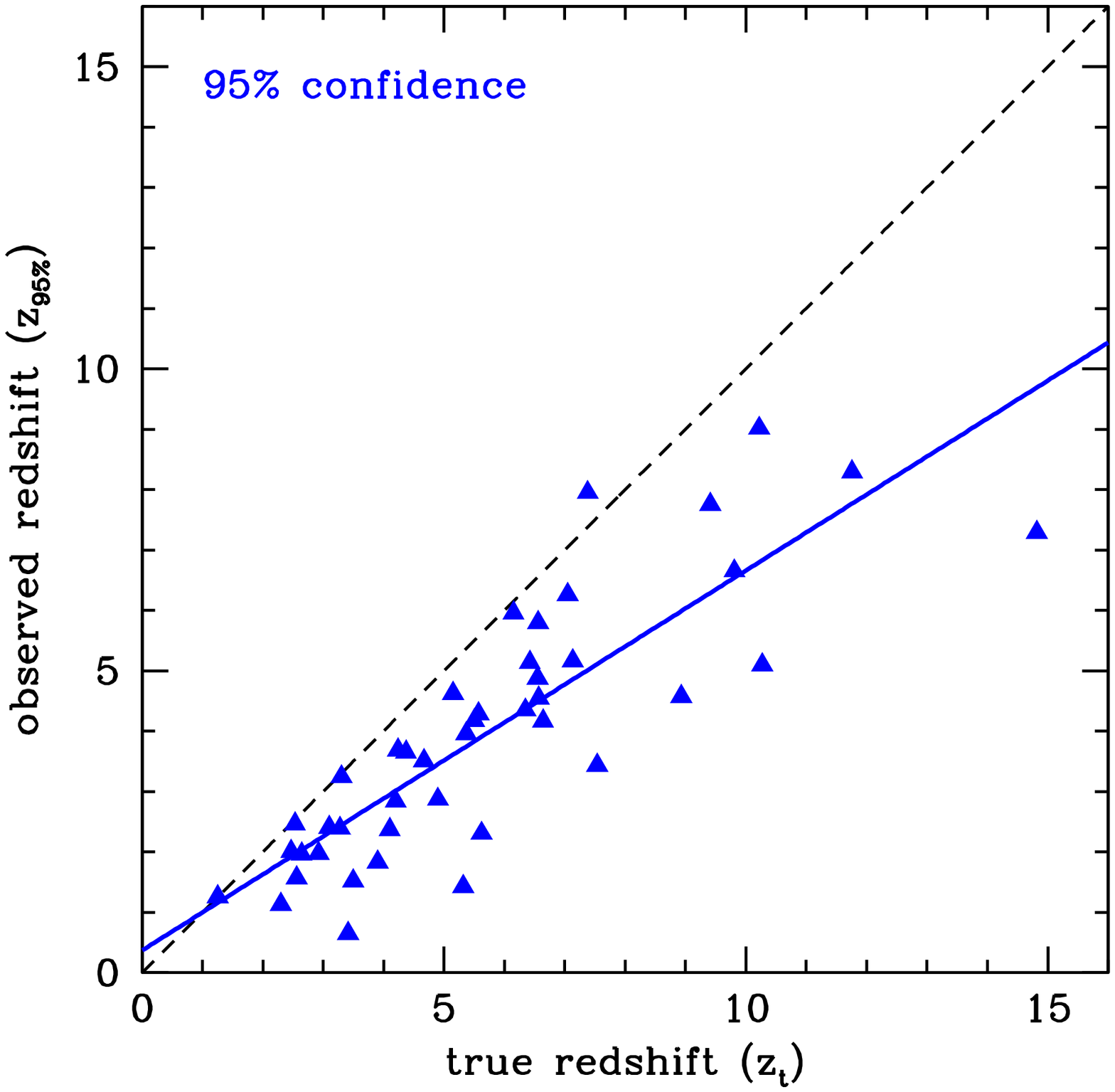}{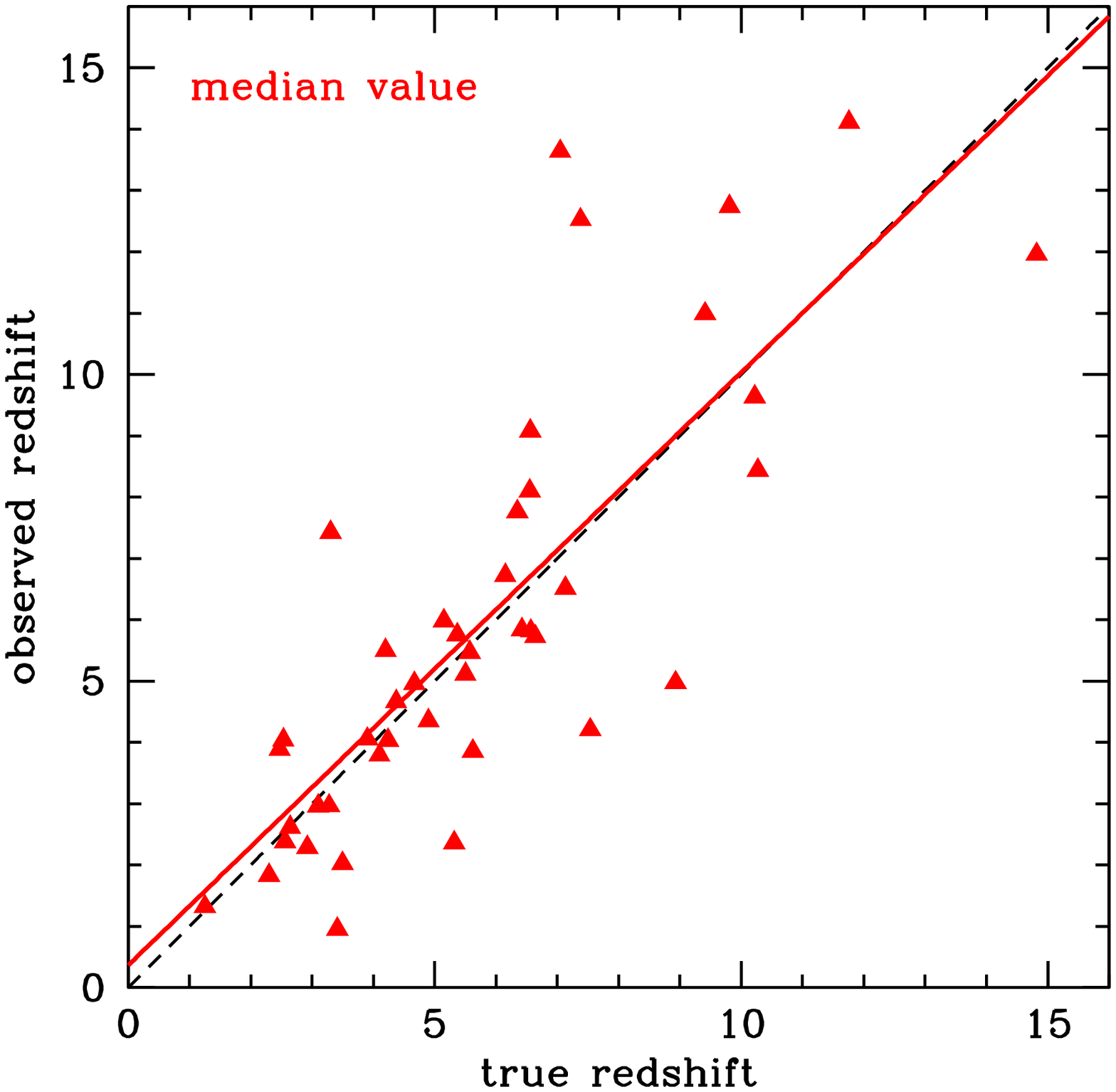}
\caption{\label{fig4}Measured versus true source redshifts. Triangles represent 43 sources for which we ran MCMCs, whereas the solid lines are the least square fits to the data sets. The left and right panels respectively show $z_{95\%}$ (see text) and the median redshift measured from the posterior distributions versus the true redshift.
}
\end{figure}

To explore our ability to go beyond the FIM approximation,
%we considered a GW signal model for the inspiral only (no merger and ring-down), but
%including sub-dominant harmonics and binary orbital precession due to some 
%misalignment between spins and the orbital angular momentum. As discussed above, the
%imprint of those features in the waveform helps to break degeneracies in the parameter space.
we chose from our meta-catalogue 43 particularly 'bad' sources for which the FIM produces errors on $D_L$ $\gtrsim100\%$ (these are usually weak sources with SNR in the range 10-20) and we ran Markov Chain Monte Carlo (MCMC) to estimate the posterior distribution of the parameters. In the vast majority of the cases we find that the posterior distribution for $D_L$ is confined to a range which is significantly smaller than the 100\% relative uncertainty predicted by the FIM. This has to be further investigated, but it is probably due to the strongly non-Gaussian posterior distribution of $D_L$ together with the fact that the  angles with which $D_L$ is partially degenerate are usually much better identified by the MCMC compared to FIM predictions. 
%Both those facts (better angle identification and non-Gaussian distribution of the posterior 
%for $D_L$) allow a much improved measurement of the luminosity distance; in other words, for 
%such 'bad' cases, the FIM largely overestimates the errors in $D_L$. 
We converted $D_L$ in $z$ assuming known cosmological parameters, and computed the value $z_{95\%}$ beyond which lies 95\% of the posterior distribution: i.e., we can say with $95\%$ confidence that a given inspiral occurred at a redshift larger than $z_{95\%}$. We also computed the median value of the posterior $z_{50\%}$ and plot the results in figure \ref{fig4}. In each panel, the 43 triangles represent the sample sources, and the solid line gives a linear least square fit for the scattered data points. The figure allows us to draw some interesting conclusions. First of all, points are quite scattered around the best fit, meaning that the lower bound we can place on the source redshift strongly depends on the individual properties of the source itself (sky location, inclination, etc). Secondly, if $z_t$ is the true redshift, the linear fits to the data imply that {\it statistically} we will be able to infer with $95\%$ confidence that a source is at a redshift larger than $z_{95\%}\approx 2/3z_t$ (blue line in the left panel of figure \ref{fig4}); for example, NGO will tell us that a source at ,e.g., $z_t=15$ lies at $z>10$ at a 95\% confidence level. Lastly, the median value of the measured source redshift $z_{50\%}$ is consistent with $z_t$ (red line in the bottom right panel), meaning that there is no systematic bias in the redshift determination. Most importantly, these exploratory tests show that NGO will be able to put strong lower limits to the redshift of those sources which are particularly faint. This, in turn, means that it will be possible to identify genuinely high redshift sources and to determine their redshift with a relative accuracy of $\lesssim 20\%$ (1$\sigma$), despite the much worse estimations based on the FIM approximation. A more comprehensiv study of the distance estimation will be presented in an upcoming paper.

\section{Reconstructing the massive black hole cosmic history through NGO observations}

NGO is an {\it observatory}. The goal is not only to detect sources, but also to extract valuable astrophysical information from the observations.  While measurements for individual systems are extremely interesting and potentially very useful for making strong-field tests of GR, it is the properties of the set of massive black hole binary mergers that are observed which will carry the most information for astrophysics. Gravitational wave observations of multiple binary mergers may be used together to learn about their formation and evolution through cosmic history.

As any observatory, NGO will observe a set of signals.  After signal extraction and data analysis, these observations will provide a catalogue of coalescing binaries, with measurements of several properties of the sources (masses, mass ratio, spins, distances, etc) and estimated errors. The interesting questions to ask are the following: \emph{can we discriminate among different massive black hole formation and evolution scenarios on the basis of gravitational wave observations alone?} Given a set of observed binary coalescences, what information can be extracted about the underlying population?  For example, will gravitational wave observations alone tell us something about the mass spectrum of the seed black holes at high redshift, that are inaccessible to conventional electromagnetic observations, or about the poorly understood physics of accretion? These questions were extensively tackled in \citep{sesana11} in the context of LISA, more technical details can be found there.

\subsection{Selection among a discrete set of models}
First we consider a discrete set of models. As argued above, in the general picture of MBH cosmic evolution, the population is shaped by the \emph{seeding process} and the \emph{accretion history}. The four models we study here are the SE, SC, LE, and LC models introduced above. As a first step, we test here if NGO observations will provide enough information to enable us to discriminate between those models, assuming that the Universe is well described by one of them.
 
Each model predicts a \emph{theoretical} distribution of coalescing MBH binaries.  A given dataset $D$ of observed events can be compared to a given model $A$ by computing the likelihood $p(D|A)$ that the observed dataset $D$ is a realisation of model $A$. When testing a dataset $D$ against a pair of models $A$ and $B$, we assign probability $p_A=p(D|A)/(p(D|A)+p(D|B))$ to model $A$, and probability $p_B=1-p_A$ to model $B$.  The probabilities $p_A$ and $p_B$ are a measure of the relative confidence we have in model $A$ and $B$, given an observation $D$.  Once NGO data is available, each model comparison will yield this single number, $p_A$, which is our confidence that model $A$ is correct. Since the NGO data set is not currently available, we can only work out how likely it is that we will achieve a certain confidence with future NGO observations.

%%%%%%%%%%%%%%%%%%%%%%%%%%%%%%%
\begin{table*}[htb]
\begin{center}
\begin{tabular}{cc}
\begin{tabular}{|l|cccc|}
\hline
\hline
\multicolumn{5}{c}{Without spins}\\ \hline
 & SE & SC & LE & LC\\
\hline
SE & $\times$ &0.48 &0.99 &0.99\\
SC & 0.53& $\times$ &1.00 &1.00\\
LE & 0.01& 0.01& $\times$ &0.79\\
LC & 0.02& 0.02& 0.22& $\times$\\
\hline
\hline
\end{tabular}&
\begin{tabular}{|l|cccc|}
\hline
\hline
\multicolumn{5}{c}{With spins}\\ \hline
 & SE & SC & LE & LC\\
\hline
SE & $\times$ &0.96 &0.99 &0.99\\
SC & 0.13& $\times$ &1.00 &1.00\\
LE & 0.01& 0.01& $\times$ &0.97\\
LC & 0.02& 0.02& 0.06& $\times$\\
\hline
\hline
\end{tabular}
\end{tabular}
\end{center}
\caption{\label{tab1}Summary of all possible comparisons of the pure models. Results are for one year of observation with NGO. We take a fixed confidence  level of $p=0.95$. The numbers in the upper-right half of each table show the fraction of realizations in which the {\it row} model will be chosen at more than this confidence level when the {\it row} model is true. The numbers in the lower-left half of each table show the fraction of realizations in which the {\it row} model {\it cannot be ruled out} at that confidence level when the {\it column} model is true. In the left table we consider the trivariate $M$, $q$, and $z$ distribution of observed events; in the right table we also include the observed distribution of remnant spins, $S_r$.}
\end{table*}
%%%%%%%%%%%%%%%%%%%%%%%%%%%%%%%

We therefore generate 1000 independent realisations of the population of coalescing MBH binaries in the Universe predicted by each of the four models.  We then simulate gravitational wave observations by producing datasets $D$ of observed events (including measurement errors), which we statistically compare to the theoretical models.  We consider only sources that are observed with SNR larger than eight in the detector. We set a confidence threshold of $0.95$, and we count what fraction of the 1000 realisations of model $A$ yield a confidence $p_A>0.95$ when compared to an alternative model $B$. We repeat this procedure for every pair of models. Here for simplicity, in modeling gravitational wave observations, we focus on circular, non-spinning binaries; therefore, each coalescing system in the population is characterised by only three intrinsic parameters --$M$, $q$, $z$ -- and we compare the \emph{theoretical} trivariate distribution $N(M,q,z)$ (i.e., we ignore spins) predicted by the models to the observed values in the dataset $D$.  In terms of gravitational waveform modeling, our analysis can therefore be considered extremely conservative. Results are shown in the left-hand panel of table \ref{tab1}, for a one year observation. The vast majority of the pair comparisons yield a $95\%$ confidence in the true model for almost all the realisations --- we can perfectly discriminate among different models. Similarly, we can always rule out the alternative (false) model at a $95\%$ confidence level. Noticeable exceptions are the comparisons of models LE to LC and SE to SC, i.e., among models differing by accretion mode only. This is because the accretion mode (efficient versus chaotic) particularly affects the spin distribution of the coalescing systems, which was not considered here. To extend this work, we added to our analysis the distribution of the merger remnant spins $S_r$ {\footnote{Because of tight time constrains, we did not use here spinning precessing waveforms, for which parameter estimation is time consuming. This can, of course, be done in the future, to exploit the full information encoded in the $N(M,q,z,a_1,a_2)$ distribution.}}, and compared the \emph{theoretical} distribution predicted by the models to the observed values (including determination errors once again). The spin of the remnant can be reasonably determined in about $30\%$ of the cases only; nevertheless, by adding this information, we are able to almost perfectly discriminate between the LE and LC and the SE and SC models, as shown in the right hand panel of table \ref{tab1}.

\subsection{Constrains on parametric models}

In the preceding section we demonstrated the potential of NGO to discriminate among a discrete set of "pure" models given a priori.  However, the true MBH population in the Universe will probably result from a mixing of known physical processes, or even from a completely unexplored physical mechanism. A meaningful way to study this problem is to construct parametric models that depend on a set of key physical parameters, $\lambda_i$, describing, for instance, the seed mass function and redshift distribution, the accretion efficiency etc. and to investigate the potential of NGO to constrain these parameters. Such a parametric family of models is not available at the moment, but we can carry out a similar exercise by mixing two of our pure models, $A$ and $B$, to produce a model in which the number of events of a particular type is given by ${\cal F}$[A]$+(1-{\cal F})[B]$, where $[A]$ is the number of events of that type predicted by model $A$, $[B]$ is the corresponding number predicted by model $B$ and ${\cal F}$ is the "mixing fraction". In this case we generate datasets $D$ from a mixed model with a certain unknown ${\cal F}$, and we estimate the ${\cal F}$ parameter by computing the likelihood that the data $D$ is drawn from a mixed distribution, as a function of ${\cal F}$. 
%A specific example is shown in figure \ref{fig.bh_7}. Here the underlying model is ${\cal F}$[SE]$+(1-{\cal F})$[LE], with ${\cal F}=0.45$. 
On a series of test cases, we found that NGO observations will allow us to pin-down the correct value of the mixing parameters with a typical uncertainty of $\sim 0.1$. Many examples of multi-model mixing are discussed in \citep{sesana11}, in the LISA context. Although highly idealised, this exercise demonstrate the potential of NGO observations to constrain the physics and astrophysics of MBHs along their entire cosmic history, in a mass and redshift range inaccessible to conventional electromagnetic observations.

\section{Conclusion} 
Future spaced based GW observation will deliver spectacular MBH science. An observatory like NGO will provide a unique survey of coalescing MBH binaries up to $z\approx 15$. Masses and spins of coalescing MBHs will be measured with unprecedented precision: individual redshifted masses will be measured with a $1-0.1\%$ relative error, luminosity distances will be recovered in some cases within a few \%, whereas individual spins will be extracted to an accuracy of 0.1-0.01. The combination of the GW observations of multiple binary mergers may be used together to learn about their formation and evolution through cosmic history, placing strong constrains on the nature of the first black hole seeds, their subsequent accretion history, and, more generally, on the early hierarchical structure formation at high redshift. The delivered science will be outstanding, unveiling the hidden, distant Universe.

%\bibliography{/Users/plagnol/Documents/physics/Donnees/Lisa/Lisa_Symposium_Paris_2012/Proceedings/myBibli/listBibli2}
%\bibliographystyle{/Users/plagnol/Documents/physics/Donnees/Lisa/Lisa_Symposium_Paris_2012/Proceedings/LateX/asp2010}

\bibliography{bibliography}
\bibliographystyle{asp2010}

 \end{document}